\documentclass[conference]{IEEEtran}
\IEEEoverridecommandlockouts
\usepackage{authblk}
\usepackage{orcidlink}
\usepackage{cite}
\usepackage{amsmath,amssymb,amsfonts}
\usepackage{algorithm}
\usepackage{algorithmic}
\usepackage{graphicx}
\usepackage{textcomp}
\usepackage{xcolor}
\usepackage{adjustbox}
\def\BibTeX{{\rm B\kern-.05em{\sc i\kern-.025em b}\kern-.08em
    T\kern-.1667em\lower.7ex\hbox{E}\kern-.125emX}}
\begin{document}

\title{Under-Sampled High-Dimensional Data Recovery via Symbiotic Multi-Prior Tensor Reconstruction}

\author{Jie Yang, Chang Su, Yuhan Zhang, Jianjun Zhu\textsuperscript{*} and Jianli Wang\textsuperscript{*} 
\thanks{The authors would like to acknowledge the support of the National Natural Science Foundation of China (Grant No. 52376159, 52474064).}
\thanks{Jie Yang,Chang Su, Yuhan Zhang and Jianli Wang are with Jiangsu Key Laboratory for Design and Manufacture of Micro-Nano Biomedical Instruments, Department of Mechanical Engineering, Southeast University, Nanjing 210096, China. Jianjun Zhu is with College of Mechanical and Transportation Engineering, China University of Petroleum, Beijing 102249, China. *Corresponding Emails: \texttt{jianjun-zhu@cup.edu.cn, wangjianli@seu.edu.cn}}
}

\maketitle

\begin{abstract}
The advancement of sensing technology has driven the widespread application of high-dimensional data. However, issues such as missing entries during acquisition and transmission negatively impact the accuracy of subsequent tasks. Tensor reconstruction aims to recover the underlying complete data from under-sampled observed data by exploring prior information in high-dimensional data. However, due to insufficient exploration, reconstruction methods still face challenges when sampling rate is extremely low. This work proposes a tensor reconstruction method integrating multiple priors to comprehensively exploit the inherent structure of the data. Specifically, the method combines learnable tensor decomposition to enforce low-rank constraints of the reconstructed data, a pre-trained convolutional neural network for smoothing and denoising, and block-matching and 3D filtering regularization to enhance the non-local similarity in the reconstructed data. An alternating direction method of the multipliers algorithm is designed to decompose the resulting optimization problem into three subproblems for efficient resolution. Extensive experiments on color images, hyperspectral images, and grayscale videos datasets demonstrate the superiority of our method in extreme cases as compared with state-of-the-art methods. 
\end{abstract}


\section{Introduction}
\label{sec:intro}
The rapid advancement of sensing technology has led to the widespread application of high-dimensional data, including color images, hyperspectral images, high-speed videos, traffic grid data, remote sensing data, and recommendation system data. However, missing values during data acquisition and transmission are often unavoidable, leading to under-sampled observations \cite{mst2}. Utilizing under-sampled data directly can significantly decrease the accuracy of subsequent tasks \cite{1}\cite{2}. Therefore, recovering complete data from under-sampled observations has become a critical and fundamental problem. 

Under-sampled high-dimensional data reconstruction is mathematically formulated as a tensor reconstruction problem. These problems belong to the ill-posed problems and possess infinite possible solutions. Currently, tensor reconstruction methods can be categorized into supervised methods based on statistical learning and unsupervised methods based on data priors. Supervised statistical learning methods rely on large-scale annotated data for training to capture contextual features and perceptual information, enabling effective reconstruction. For instance, He et al. \cite{3} introduced the Masked Auto-Encoder (MAE), which compels the model to infer masked elements using masked reconstruction applied to extensive training data. However, such methods depend heavily on training datasets and often lack interpretability. The second category leverages the inherent prior information in high-dimensional data to address tensor reconstruction problem through iterative optimization. High-dimensional data priors can be classified into global correlation, local smoothness, and non-local similarity. Global correlation is mathematically represented as low rank. Liu et al. \cite{4} pioneered using low-rank as prior constraints, achieving notable results in tensor reconstruction. They defined tensor rank via linear transformation decomposition. Subsequently, Luo et al. \cite{5} proposed tensor decompositions based on nonlinear transformations and tensor function representations \cite{6} to define tensor rank. Beyond low-rank priors, high-dimensional data often exhibit similar elements in adjacent areas and similar blocks across different areas, mathematically represented as local smoothness and non-local similarity, respectively. Local smoothness preserves local details by employing regularization on adjacent areas. Representative techniques include Total Variation (TV) \cite{7}, framelet \cite{8}, and CNN \cite{9}. Local smoothness is often combined with low-rank prior. For example, Li et al. \cite{10} combined the tensor Tucker rank prior with TV regularization, resulting in significant improvements in reconstruction performance. Non-local similarity is primarily implemented through filtering processes on stacked tensor sub-blocks, with representative methods including non-local mean filtering \cite{11}, block-matching 3D filtering (BM3D) \cite{12}, and block-matching 4D filtering (BM4D) \cite{13}.

The above methods have made promising results in under-sampled data reconstruction tasks for color images \cite{12}, hyperspectral images\cite{14}, videos \cite{13}, traffic grid data\cite{15}, remote sensing data \cite{16}, health monitoring data \cite{mst1} and recommendation system data \cite{17}. However, these methods still suffer from poor performance under difficult conditions where sampling rate is extremely low. To explore potential solutions, we analyze the priors used in the literature. The comparison reveals that combining two types of priors improves performance compared to using only one. This leads us to hypothesize that the poor performance in extreme cases stems from the failure to fully exploit prior information in the data. Intuitively, incorporating three types of priors may further enhance performance.

This paper proposes a framework combining three types of priors—low-rankness, local smoothness, and non-local similarity—to address the challenges of under-sampled data reconstruction. To adapt to the high-dimensional data in real-world scenarios, a non-linear learnable tensor decomposition \cite{5} is used to enforce the tensor low rank constraints on the reconstructed data. Meanwhile, a CNN \cite{27} trained on large-scale datasets is employed as a local smoothing regularization. Additionally, the BM3D \cite{12} algorithm is incorporated to enhance the non-local similarity of the reconstructed data. We formulate a constrained optimization problem that incorporates observed data as hard constraints while jointly considering all three priors. The ADMM framework \cite{18} is used to decompose the resulting problem into subproblems, enabling efficient solutions. The effectiveness of the proposed algorithm is validated on color images, hyperspectral images, and grayscale video datasets. The main contributions of this work are summarized as follows:
\begin{itemize}
  \item A novel multi-prior tensor reconstruction framework is proposed. A learnable tensor decomposition is used to constrain the tensor rank. A pre-trained CNN is employed as a local smoothing regularization, and the BM3D algorithm is incorporated to enhance the non-local similarity.
  \item An ADMM-based solution algorithm is designed, which decomposes the resulting constrained optimization problem into three subproblems for efficient solutions.
  \item Extensive experiments are conducted, which validate the effectiveness of the proposed model and algorithm. Results show that the proposed approach achieves competitive performance on color images, hyperspectral images, and grayscale video datasets.
\end{itemize}

The outline of this paper is organized as below. Section \uppercase\expandafter{\romannumeral2} reviews related work on tensor reconstruction. Section \uppercase\expandafter{\romannumeral3} presents the necessary preliminary theories and the proposed algorithm. Section \uppercase\expandafter{\romannumeral4} reports the comparative experimental results and ablation studies to demonstrate the performance of the proposed method. 

\section{Related Work}
\textbf{One Prior} Tensor reconstruction can be achieved by independently using one prior, either the low-rank property of high-dimensional data or non-local similarity. Tensor rank has multiple definitions \cite{19}, such as CANDECOMP/PARAFAC (CP) rank, Tucker rank, Tubal rank, Tensor Train rank, and Tensor ring rank. Different tensor reconstruction methods have been proposed based on different tensor rank definitions. Zhao et al. \cite{n35} applied adaptive CP rank minimization to high-dimensional data reconstruction. Minimizing the Tucker rank constraint is NP-hard. Liu et al. \cite{4} proposed using the nuclear norm minimization of unfolded matrices as the convex relaxation of Turker rank minimization. Non-convex relaxations, such as the logarithmic nuclear \cite{n36} norm and Schatten-p nuclear norm \cite{n37}, were introduced and achieved excellent results in tensor reconstruction. Minimizing the Tubal rank constraint is also NP-hard. Semerci et al. \cite{20} proposed the tensor nuclear norm (TNN) as the convex relaxation of Tubal rank minimization. Zhang et al. \cite{n39} applied TNN minimization to tensor reconstruction, while non-convex relaxations, such as log-TNN \cite{n40} and Laplace-TNN \cite{n41}, were also proposed and applied to tensor reconstruction. To reduce the computational complexity of tensor Singular Vector Decomposition (t-SVD) decomposition, tensor-tensor product methods were proposed to accelerate the computation process\cite{n42}\cite{n43}. Bengua et al.\cite{21} applied Tensor Train rank minimization to color image and video reconstruction. Yuan et al. \cite{22} used Tensor Ring rank minimization for color image reconstruction. These methods achieved tensor reconstruction using only low-rank priors, while non-local similarity priors alone yielded some results, the use of non-local similarity priors alone has also achieved some results. For instance, the Kronecker-Basis-Representation (KBR) algorithm \cite{23} utilized non-local similarity for data reconstruction. BM3D \cite{12} and BM4D \cite{13} were applied to image and video denoising, respectively, achieving remarkable results.

\textbf{Two Priors} Combining various priors can further enhance reconstruction performance. Qiu et al. \cite{24} integrated TNN with TV regularization for hyperspectral image and video reconstruction.  Ding et al.\cite{n44} combined Tensor Train rank with TV regularization to suppress block artifacts in reconstructed data. He et al. \cite{n23} effectively reconstructed hyperspectral images by integrating Tensor Ring rank and TV regularization. Jiang et al. \cite{25} replaced TV regularization with framelet, while Zhang \cite{26} and Zhao et al. \cite{9} proposed using pre-trained CNN \cite{27} instead of TV regularization.  Ji et al. \cite{8} applied a method combining Tucker rank and non-local similarity regularization to reconstruct remote sensing images. Ding et al. combined Tensor Train rank with non-local similarity and demonstrated excellent performance in natural image reconstruction tasks. Chen et al. \cite{n24} used Tensor Ring rank combined with non-local similarity for hyperspectral image denoising. Zhao et al. \cite{28} explored integrating Tensor Train rank constraints with TV smoothing and non-local priors.

\begin{figure}[htbp]
\centering
\includegraphics[width=0.5\textwidth]{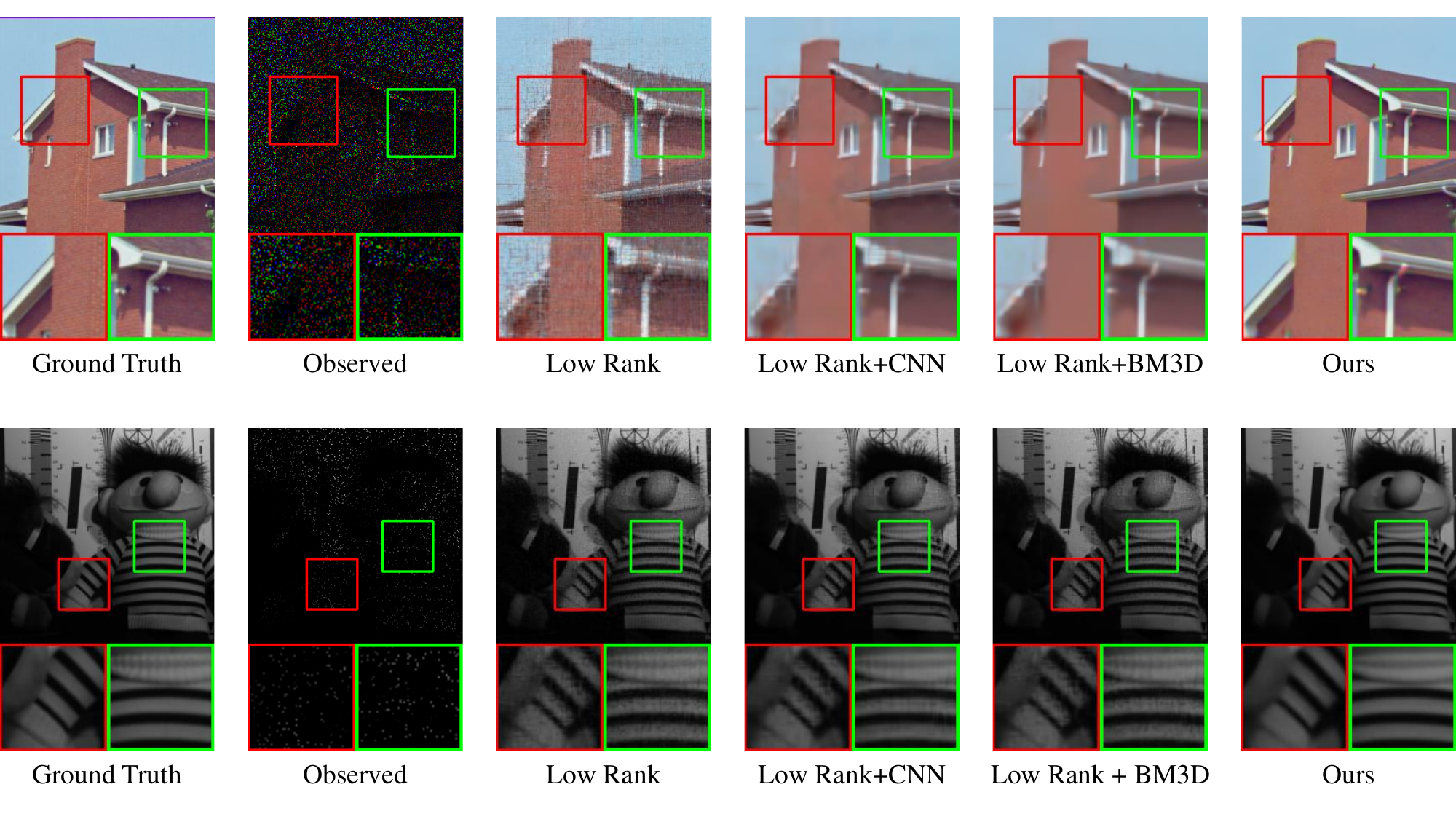}
\caption{The three types of prior are symbiotic with each other.}
\label{fig1}
\end{figure}

Inspired by previous work, we propose a novel tensor reconstruction method leveraging multiple priors. Figure \ref{fig1} visually demonstrates the contributions of each prior to the reconstruction process. The results from each component highlight how these priors work together synergistically. The limitations of one prior are compensated by the strengths of the others, resulting in superior reconstruction quality, even under extreme under-sampling conditions.

\section{Methodology}

This section presents a tensor reconstruction framework that integrates three complementary priors: low-rank structure through learnable tensor decomposition, local smoothness via pre-trained CNN regularization, and non-local similarity through BM3D filtering. These priors work together to effectively recover missing entries by formulating reconstruction as a constrained optimization problem that incorporates observed data as hard constraints. An ADMM-based algorithm decomposes this problem into efficiently solvable subproblems that are solved iteratively to obtain the final reconstructed tensor.

\subsection{Multi-Prior Tensor Reconstruction Model}
To formulate a tensor reconstruction framework that leverages multiple complementary priors, we propose an optimization problem that integrates low-rank structure, local smoothness, and non-local similarity, which is expressed by:
\begin{equation}
\begin{split}
& \arg\min_{X, A, B} \,\, \Vert P_{\Omega}(X - O)\Vert^2_F+\alpha PTV(A, B) \\
& \quad \quad \quad \quad + \beta R_L(X) + \gamma R_N(X) \\
&s.t.\quad  
X=g(A \Delta B)
\end{split}
\end{equation}
Here, \emph{$\Vert P_{\Omega}(X-O)\Vert^2_F$} is the fidelity term used to constrain the difference between the reconstructed data \emph{$X$} and the observed undersampled data \emph{$O$}. $P_{\Omega}(\cdot)$ is projection function that keeps the entries in $\Omega$ while sets others be zeros. \emph{$g(A\Delta B)$} represents the learnable tensor decomposition, which enforces the low rank of the reconstructed data \emph{$X$}. The tensor \emph{$X \in R^{H \times W \times C}$} is decomposed into the tensor product of two smaller-shaped factor tensors \emph{$A \in \mathbb{R}^{H \times r \times C}$} and \emph{$B \in \mathbb{R}^{r \times W \times C}$}, where \emph{$\Delta$} represents the tensor-tensor product \cite{t20} and \emph{$g(\cdot)$} is the inverse operation of tensor decomposition operation. \emph{$PTV(\cdot)$} is the factor-tensor gradient reularization. 
\begin{equation}
PTV(A, B) = \Vert \nabla A_x \Vert_{l_1}+\Vert \nabla B_y \Vert_{l_1}
\end{equation}

\emph{$R_L(\cdot)$} is the local smoothness regularization implemented using a pre-trained CNN network. \emph{$R_N(\cdot)$} is the non-local similarity regularization implemented using the BM3D algorithm. \emph{$\alpha$}, \emph{$\beta$}, and \emph{$\gamma$} are weight coefficients. The coupling of three different properties of the prior makes the objective function difficult to optimize directly. Therefore, the auxiliary variable \emph{Y} is introduced to decouple local and non-local priors term, then we have:
\begin{equation}
\begin{split}
& \arg\min_{X, Y, A, B} \,\,\Vert P_{\Omega}(X - O)\Vert^2_F+\alpha PTV(A, B) \\
& \quad \quad \quad \quad + \beta R_L(X) + \gamma R_N(Y) \\
&s.t.\quad  
X=g(A \Delta B), \,\, X=Y
\end{split}
\label{eq3}
\end{equation}

\subsection{ADMM Solver}
The Lagrange augmented function of Eq.(3) is
\begin{equation}
\begin{split}
&L(X, Y, A, B, M_1, M_2)=  \Vert P_{\Omega}(X - O)\Vert^2_F+\alpha PTV(A, B)  \\
& + \beta R_L(X) + \gamma R_N(Y) +  \langle X-g(A\Delta B), M_1\rangle\\
&  +\langle X-Y, M_2 \rangle+\frac{\rho}{2}\Vert X-g(A\Delta B) \Vert_F^2+\frac{\psi}{2}\Vert X-Y \Vert_F^2 \\
\end{split}
\end{equation}

According to the ADMM framework, Eq. (4) is decomposed into multiple subproblems that are iteratively solved and updated in an alternating manner:

\begin{equation}
\left\{\begin{array}{lc}
A^{(s+1)},B^{(s+1)}\\
\quad=\mathop{\arg\min}\limits_{A, B} L[X^{(s)}, Y^{(s)}, A, B, M^{(s)}_1, M^{(s)}_2]\\
X^{(s+1)}\\
\quad=\mathop{\arg\min}\limits_{X} L[X, Y^{(s)}, A^{(s+1)}, B^{(s+1)}, M^{(s)}_1, M^{(s)}_2]\\
Y^{(s+1)}\\
\quad=\mathop{\arg\min}\limits_{Y} L[X^{(s+1)}, Y, A^{(s+1)}, B^{(s+1)}, M^{(s)}_1, M^{(s)}_2]\\
M_1^{(s+1)} = M_1^{(s)} + \rho\left(X^{(s+1)}-g(A\Delta B)^{(s+1)}\right) \\
M_2^{(s+1)} = M_2^{(s)} + \psi\left(X^{(s+1)}-Y^{(s+1)}\right) \\
\end{array}\right.
\end{equation}
Next, we discuss the details for solving the subproblems.

\textbf{Subproblem 1}: update \emph{$A$}, \emph{$B$}
\begin{equation}
\begin{split}
&A^{(s+1)}, B^{(s+1)} = \arg\min_{A, B}  \Vert P_{\Omega}(g(A\Delta B) - O)\Vert^2_F  \\
& +\alpha PTV(A, B) +  \langle X^{(s)}-g(A\Delta B), M^{(s)}_1\rangle \\
& +\frac{\rho}{2}\Vert X^{(s)}-g(A\Delta B) \Vert_F^2 \overset{def}{=}  \arg\min_{A, B} J(A, B)\\
\end{split}
\label{eq6}
\end{equation}
This subproblem can be viewed as a parameter optimization problem for a neural network by simply computing the loss function and updating the parameters using gradient backpropagation and gradient descent optimizer.
\begin{equation}
\left\{\begin{array}{lc}
A^{(s+1)} = A^{(s)} - \alpha \frac{\partial {J(A, B)}}{\partial{A}}\\
A^{(s+1)} = A^{(s)} - \alpha \frac{\partial {J(A, B)}}{\partial{B}}\\
\end{array}\right.
\label{eq7}
\end{equation}
here, $\alpha$ is learning rate.

\textbf{Subproblem 2}: update \emph{$X$}
\begin{equation}
\begin{split}
&X^{(s+1)}=\arg\min_{X} \, \beta R_L(X)  + \langle X-g(A\Delta B)^{(s+1)}, M^{(s)}_1\rangle  \\
&  +\langle X-Y^{(s)}, M^{(s)}_2 \rangle +\frac{\rho}{2}\Vert X-g(A\Delta B)^{(s+1)} \Vert_F^2\\
& +\frac{\psi}{2}\Vert X-Y^{(s)} \Vert_F^2 \\
\end{split}
\end{equation}
Let $\sigma_1 = \sqrt{\frac{\beta}{\rho+\psi}}$, the close-form solution of subproblem 2 is 
\begin{equation}
\begin{split}
&X^{(s+1)}= \\
&R_L\left[\frac{\rho (g(A\Delta B)^{(s+1)}-\frac{M^{(s)}_1}{\rho})+\psi(Y^{(s)}-\frac{M^{(s)}_2}{\psi})}{\rho+\psi}, \sigma_1\right]
\end{split}
\label{eq9}
\end{equation}

\textbf{Subproblem 3}: update \emph{$Y$}
\begin{equation}
\begin{split}
Y^{(s+1)} = &\arg\min_{Y} \gamma R_N(Y) +\langle X^{(s+1)}-Y, M^{(s)}_2 \rangle \\
& +\frac{\psi}{2}\Vert X^{(s+1)}-Y \Vert_F^2 \\
\end{split}
\label{eq10}
\end{equation}

Let $\sigma_2 = \sqrt{\frac{\gamma}{\psi}}$, the close-form solution of subproblem 3 is
\begin{equation}
Y^{(s+1)}= R_N \left[ X^{(s+1)}+\frac{M^{(s)}_2}{\psi}, \sigma_2\right]
\label{eq11}
\end{equation}

Update Lagrange multipliers $M_1$ and $M_2$, then we have
\begin{equation}
\left\{\begin{array}{lc}
M_1^{(s+1)} = M_1^{(s)} + \rho\left(X^{(s+1)}-g(A\Delta B)^{(s+1)}\right) \\
M_2^{(s+1)} = M_2^{(s)} + \psi\left(X^{(s+1)}-Y^{(s+1)}\right)\\
\end{array}\right.
\label{eq12}
\end{equation}

Since subproblem 1 contains more variables and parameters than the other subproblems, an inner loop structure is designed to ensure that all the variables can be updated synchronously. The complete algorithm is as follows.

\begin{algorithm}[!h]
    \caption{The ADMM Algorithm for Solving Eq. (\ref{eq3})}
    \label{alg:AOA}
    \renewcommand{\algorithmicrequire}{\textbf{Input:}}
    \renewcommand{\algorithmicensure}{\textbf{Initialization:}}
    \begin{algorithmic}[1]
        \REQUIRE The observed data $O$, the number of external loops $K$, the number of internal loops $L$, the threshold of relative rate of change $\epsilon$.  
        \ENSURE $A, B$, $X=Y=O$, $M_1=M_2=0$.    
        \WHILE{$i \le K$ and $(X^{i}-X^{i-1}) / X^{i-1}>\epsilon$}
        
            \FOR{each $j \in [1, L] $}
                \STATE update $A$, $B$ by Eq. (\ref{eq7})
            \ENDFOR
        
        \STATE update $X$ by Eq.(\ref{eq9})
        
        \STATE update $Y$ by Eq. (\ref{eq11})
        
        \STATE update $M_1$, $M_2$ by Eq. (\ref{eq12})
        
        \ENDWHILE
        
        \RETURN $X$, $Y$, $g(A \Delta B)$
    \end{algorithmic}
\textbf{Output:} The reconstructed tensor $X$. 
\end{algorithm}

\section{Results and Discussions}
In this section, we present comprehensive experimental results to validate the effectiveness of our proposed tensor reconstruction framework. We first describe the implementation details, including datasets, evaluation metrics, and experimental settings. Then, we conduct extensive comparison experiments on color images, hyperspectral images, and grayscale video datasets to demonstrate the superior performance of our method. We also perform ablation studies to analyze the contribution of each prior component. Finally, we provide detailed discussions on reconstruction quality analysis, hyperparameter effects, and convergence behavior.

\subsection{Implementation Details}\label{AA}
In this study, experiments are conducted on various datasets, including color images \footnote{http://sipi.usc.edu/database/database.php}, hyperspectral images \footnote{http://www.cs.columbia.edu/CAVE/databases/multispectral/}, and grayscale videos \footnote{http://trace.eas.asu.edu/yuv/}. Detailed information about these public datasets is provided in Table \ref{tab_information}. The process for constructing under-sampled observed data \emph{O}, is described as follows. First, a MASK tensor with the same shape as the data sample \emph{X} is created, with each element of the MASK initialized to ones. Next, some of the elements in the MASK are randomly replaced with zeros, representing the absence of data entries. The data sample \emph{X} is then element-wise multiplied by the modified MASK to obtain the under-sampled observation data \emph{O}. The ratio of non-zero elements in \emph{O} to the total number of elements is defined as the sampling rate (SR). For instance, when 80$\%$ of the elements in $X$ are replaced with zeros, the sampling rate becomes 20$\%$.  Following random masking, the proportion of remaining data ranges from 30$\%$ \text{to} 1$\%$, corresponding to a sampling rate of 30$\%$\text{to} 1$\%$.

\begin{table}[htbp]
\caption{Detailed Information about datasets}
\begin{center}
\begin{adjustbox}{width=0.5\textwidth}
\begin{tabular}{cccc}
\hline
 & \textbf{Color Images} & \textbf{Hyperspectral Images} & \textbf{Grayscale Videos}\\
\hline
Data Shape& $256\times256\times3$ &$256\times256\times30$ & $256\times256\times3$ \\
Upper Bound & 255 & 65535 & 256 \\
Statistics & 4 & 8 & 5 \\
Names & \emph{Baboon}, \emph{House} et al. & \emph{Balloons}, \emph{Toy} et al. & \emph{Suzie} et al. \\
\hline
\end{tabular}
\label{tab_information}
\end{adjustbox}
\end{center}
\end{table}

\begin{table*}[htbp] 
\caption{Comparison of the PSNR of reconstructed data from different methods}
\begin{center}
\begin{adjustbox}{width=\textwidth}
\begin{tabular}{ccccccc|cccccc|cccccc}
\hline
 & \multicolumn{6}{c}{\textbf{Color Image \emph{Baboon}}} & \multicolumn{6}{c}{\textbf{Hyperspectral Image \emph{Balloons}}} & \multicolumn{6}{c}{\textbf{Grayscale Video \emph{Suzie}} } \\
\cline{2-19} 
\textbf{SR} 
& \textbf{\textit{1\%}} & \textbf{\textit{3\%}} & \textbf{\textit{5\%}} & \textbf{\textit{10\%}} & \textbf{\textit{20\%}} & \textbf{\textit{30\%}} & \textbf{\textit{1\%}} & \textbf{\textit{3\%}} & \textbf{\textit{5\%}} & \textbf{\textit{10\%}} & \textbf{\textit{20\%}} & \textbf{\textit{30\%}} &\textbf{\textit{1\%}} & \textbf{\textit{3\%}} & \textbf{\textit{5\%}}& \textbf{\textit{10\%}}& \textbf{\textit{20\%}} & \textbf{\textit{30\%}} \\
\hline
Observed&5.43&5.52&5.61&5.84&6.36&6.94&11.67&11.76&11.85&12.09&12.60&13.18&6.88&6.97&7.06&7.30&7.81&8.39\\
MCALM \cite{t5}&7.08&14.45&16.21&17.74&19.73&21.22&13.30&20.59&25.18&30.14&35.10&38.36&9.55&15.65&18.65&21.38&24.82&27.24 \\
HaLRTC \cite{4}&6.18&12.88&15.11&17.67&20.10&21.72&12.84&20.78&27.41&33.34&39.25&43.18&9.62&16.78&19.69&23.32&27.04&29.52\\
Tmac \cite{t7}&5.51&5.59&5.68&5.92&6.44&7.01&11.43&11.54&11.67&12.05&12.84&13.95&7.92&8.05&8.16&8.60&9.58&10.80\\
TraceTV \cite{t8}&5.54&5.80&6.23&9.78&16.02&18.75&11.45&12.78&17.50&33.85&42.32&46.45&7.93&10.02&19.48&26.81&30.81&33.43\\
t-SVD \cite{t9}&12.23&14.24&15.58&17.52&19.74&21.49&23.66&29.14&32.14&36.85&42.17&45.86&20.65&23.64&25.26&27.62&30.46&32.75\\
McpTC \cite{t10}&8.33&15.61&16.60&17.97&20.11&21.72&24.22&33.21&35.49&39.70&44.96&47.91&18.03&23.75&25.62&38.35&31.22&33.33\\
ScadTC \cite{t10}&7.28&14.69&15.63&16.95&19.37&21.03&23.76&34.00&35.76&39.79&44.90&47.77&18.81&24.64&26.09&28.55&31.28&33.37\\
KBR \cite{14} &13.82&15.33&17.21&18.69&19.83&20.98&20.14&34.85&40.44&46.31&\textbf{50.97}&\textbf{52.73}&19.54&24.96&28.02&31.45&34.76&\textbf{37.33}\\
HLRTF \cite{5} &12.96&14.12&16.33&18.46&20.98&22.58&27.51&34.44&40.51&45.26&48.29&51.55&23.76&26.09&28.28&31.51&34.67&36.83\\ 
LRTFR \cite{6} &11.51&14.46&18.18&19.20&20.09&20.51&29.88&33.32&34.61&36.54&38.01&39.34&24.56&25.54&26.93&27.72&28.60&29.05\\
Ours &\textbf{14.69}&\textbf{19.32}&\textbf{20.04}&\textbf{21.28}&\textbf{22.87}&\textbf{24.04}&\textbf{36.24}&\textbf{42.28}&\textbf{44.38}&\textbf{47.39}&50.14&51.68&\textbf{26.34}&\textbf{29.48}&\textbf{30.87}&\textbf{32.68}&\textbf{36.97}&36.61\\
\hline
\end{tabular}
\label{tab_psnr}
\end{adjustbox}
\end{center}
\end{table*}

\begin{table*}[htbp] 
\caption{Comparison of the SSIM of reconstructed data from different methods}
\begin{center}
\begin{adjustbox}{width=\textwidth}
\begin{tabular}{ccccccc|cccccc|cccccc}
\hline
 & \multicolumn{6}{c}{\textbf{Color Image \emph{Baboon}}} & \multicolumn{6}{c}{\textbf{Hyperspectral Image \emph{Balloons}}} & \multicolumn{6}{c}{\textbf{Grayscale Video \emph{Suzie}} } \\
\cline{2-19} 
\textbf{SR} 
& \textbf{\textit{1\%}} & \textbf{\textit{3\%}} & \textbf{\textit{5\%}} & \textbf{\textit{10\%}} & \textbf{\textit{20\%}} & \textbf{\textit{30\%}} & \textbf{\textit{1\%}} & \textbf{\textit{3\%}} & \textbf{\textit{5\%}} & \textbf{\textit{10\%}} & \textbf{\textit{20\%}} & \textbf{\textit{30\%}} &\textbf{\textit{1\%}} & \textbf{\textit{3\%}} & \textbf{\textit{5\%}}& \textbf{\textit{10\%}}& \textbf{\textit{20\%}} & \textbf{\textit{30\%}} \\
\hline
Observed &0.002&0.006&0.009&0.018&0.036&0.057&0.040&0.057&0.066&0.089&0.124&0.151&0.003&0.007&0.009&0.013&0.020&0.027\\
MCALM \cite{t5}&0.030&0.120&0.172&0.252&0.398&0.524&0.234&0.522&0.719&0.868&0.946&0.979&0.061&0.198&0.355&0.520&0.704&0.807\\
HaLRTC \cite{4}&0.029&0.143&0.188&0.285&0.440&0.568&0.377&0.751&0.878&0.945&0.980&0.992&0.182&0.463&0.555&0.680&0.803&0.870\\
Tmac \cite{t7}&0.002&0.006&0.009&0.018&0.037&0.059&0.048&0.060&0.065&0.080&0.096&0.343&0.006&0.012&0.016&0.025&0.044&0.071\\
TraceTV \cite{t8}&0.004&0.014&0.030&0.149&0.414&0.558&0.060&0.357&0.674&0.952&0.986&0.993&0.008&0.163&0.586&0.804&0.899&0.938\\
t-SVD \cite{t9}&0.074&0.117&0.157&0.250&0.408&0.550&0.582&0.797&0.874&0.946&0.980&0.991&0.474&0.618&0.689&0.779&0.863&0.909\\
McpTC \cite{t10}&0.026&0.110&0.152&0.246&0.416&0.558&0.708&0.937&0.952&0.978&0.991&0.995&0.421&0.654&0.722&0.809&0.885&0.923\\
ScadTC \cite{t10}&0.019&0.084&0.124&0.205&0.382&0.524&0.609&0.942&0.954&0.977&0.991&0.995&0.420&0.666&0.727&0.811&0.885&0.923\\
KBR \cite{14} &0.100&0.137&0.195&0.287&0.403&0.515&0.562&0.937&0.979&0.992&0.995&0.997&0.472&0.682&0.796&\textbf{0.883}&0.936&0.962\\
HLRTF \cite{5} &0.087&0.155&0.233&0.341&0.508&0.623&0.731&0.909&0.977&0.990&0.995&0.997&0.540&0.640&0.751&0.865&0.923&0.943\\
LRTFR \cite{6} &0.058&0.138&0.272&0.294&0.320&0.337&0.908&0.939&0.947&0.961&0.967&0.974&0.688&0.721&0.749&0.767&0.789&0.802\\
Ours &\textbf{0.212}&\textbf{0.357}&\textbf{0.417}&\textbf{0.512}&\textbf{0.640}&\textbf{0.713}&\textbf{0.973}&\textbf{0.988}&\textbf{0.991}&\textbf{0.994}&\textbf{0.996}&\textbf{0.997}&\textbf{0.722}&\textbf{0.812}&\textbf{0.844}&0.878&\textbf{0.946}&\textbf{0.969}\\
\hline
\end{tabular}
\label{tab_ssim}
\end{adjustbox}
\end{center}
\end{table*}

\begin{figure*}[htbp]
\centering
\includegraphics[width=\textwidth]{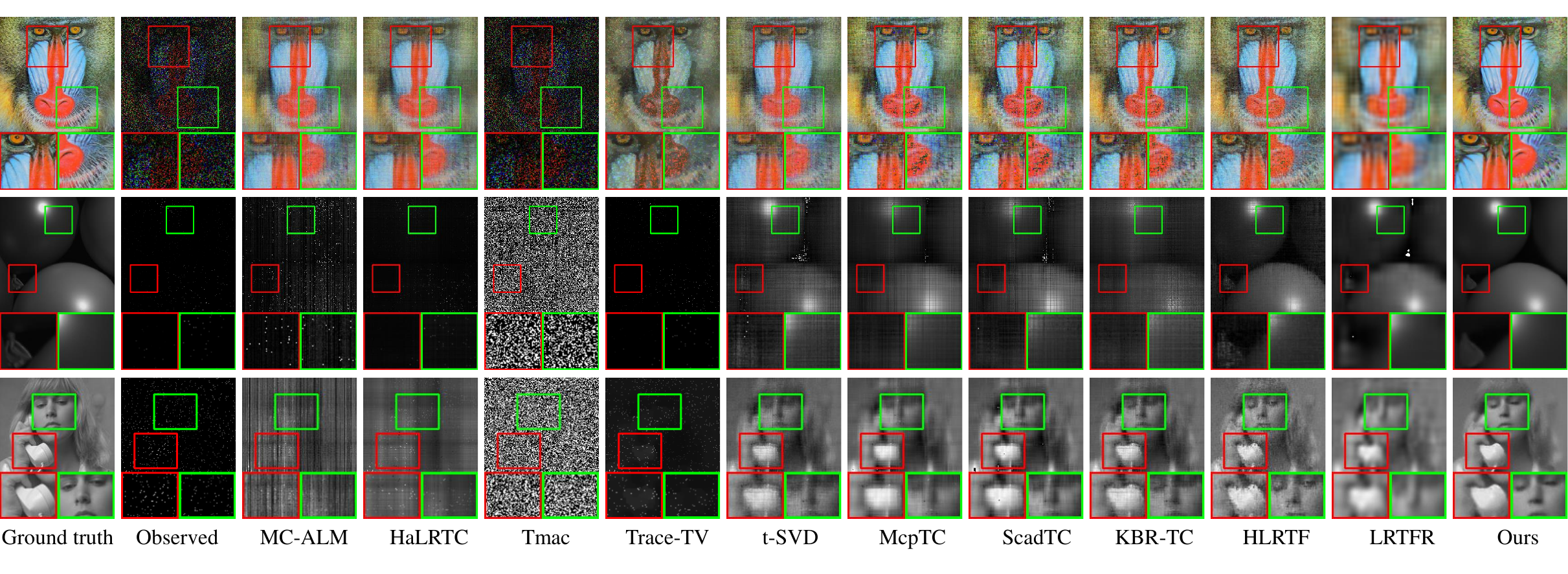}
\caption{Comparison of reconstruction details in the reconstructed data using different methods for color image \emph{Baboon} (first row, SR=20\%), hyperspectral image \emph{Balloons} (second row, SR=3\%) and grayscale video \emph{Suzie} (third row, SR=1\%)}
\label{fig2}
\end{figure*}

The performance of different reconstruction methods is quantified using two metrics, Peak signal-to-noise Ratio (PSNR) and Structural similarity (SSIM). Higher values of these two metrics indicate better performance of the reconstruction algorithm. This study sets the number of internal iterations to 15, the number of external iterations to 100, and the relative change rate threshold to 0.01. The discussion on the smoothing intensity parameter can be found in \emph{C.Discussions}. The hyperparameters of the comparison methods are adjusted based on the recommendations in the respective papers and corresponding codes to ensure optimal performance.

\subsection{Comparison Experiments}
In this section, we evaluate our proposed method through extensive comparative experiments on the color images dataset, hyperspectral images dataset, and grayscale videos dataset. Tables \ref{tab_psnr} and \ref{tab_ssim} compare the reconstruction quality metrics of various methods on the three types of datasets. The results in Table \ref{tab_psnr} and \ref{tab_ssim} are from a single example for each dataset type, with additional examples provided in the \emph{Supplementary Material}. 
At identical sampling rates, the proposed method achieves a higher PSNR and SSIM across all three datasets, indicating its clear superiority over other methods. Within the same dataset, reconstruction quality improves for all methods as the sampling rate increases, owing to more observed data. However, under extreme under-sampling rates of only 1\% \text{to} 5\%, some methods fail due to insufficient observed data. In contrast, the proposed algorithm still outperforms others significantly in these extreme scenarios, demonstrating its robust reconstruction capability even with minimal observed data. Compared to a hyperspectral image and grayscale video, a color image has only three channels in the third dimension, which makes reconstruction more challenging and results in lower quality metrics at the same sampling rate.

\subsection{Discussions}

Figure \ref{fig2} compares the reconstruction details of the Baboon (SR=20\%), Balloons (SR=1\%), and Suzie (SR=3\%) datasets under different methods. When the observed data is highly sparse, reconstructions based solely on low-rank properties appear blurry and exhibit concentrated regions of failure, which visually manifest as noise. The results demonstrate that the proposed method effectively mitigates the blurriness and noise issues seen in other methods, delivering significantly better visual performance.

\begin{figure}[htbp]
\centering
\includegraphics[width=0.5\textwidth]{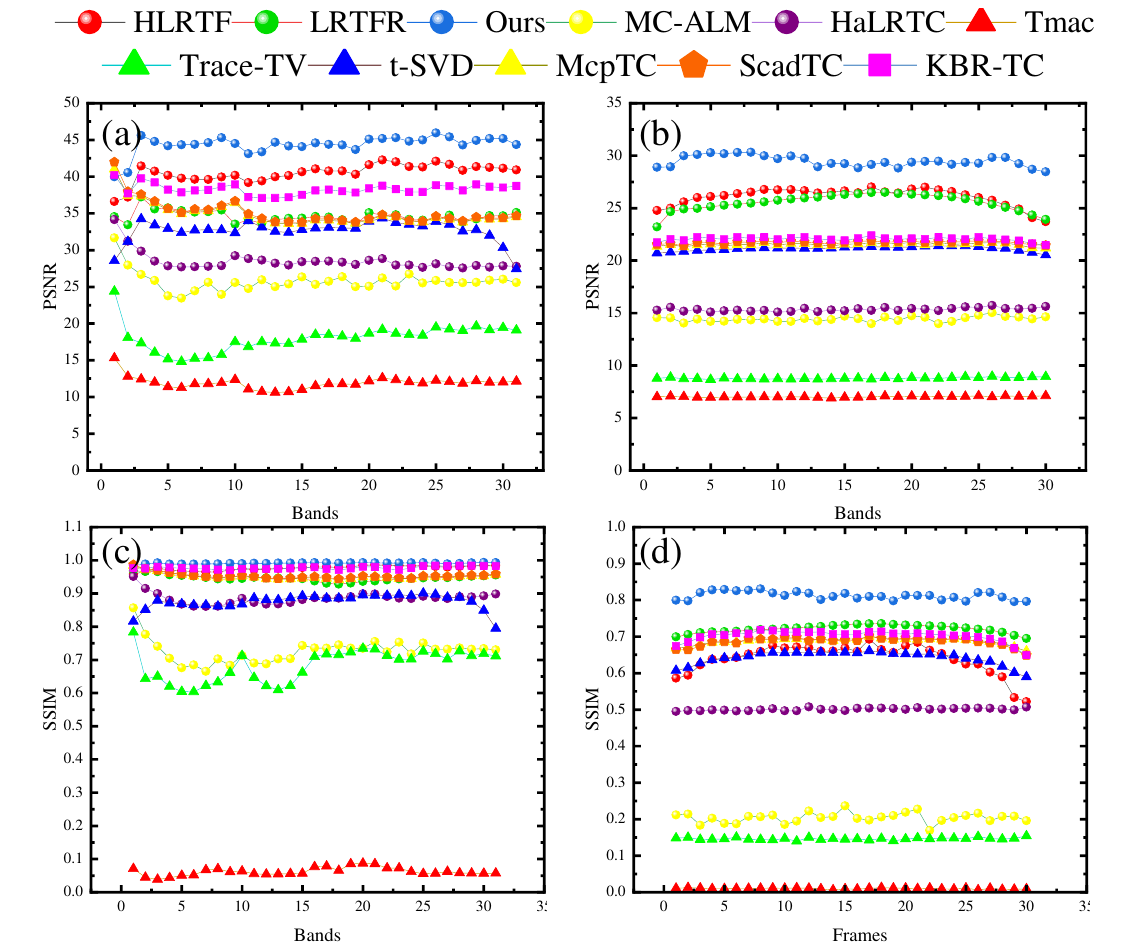}
\caption{ PSNR and SSIM values with respect to the frame/band number of the
recovered (a)(c) hyperspectral image \emph{Balloons} and (b)(d) grayscale video \emph{Suzie} by different
methods.}
\label{fig3}
\end{figure}

\begin{figure}[htbp]
\centering
\includegraphics[width=0.45\textwidth]{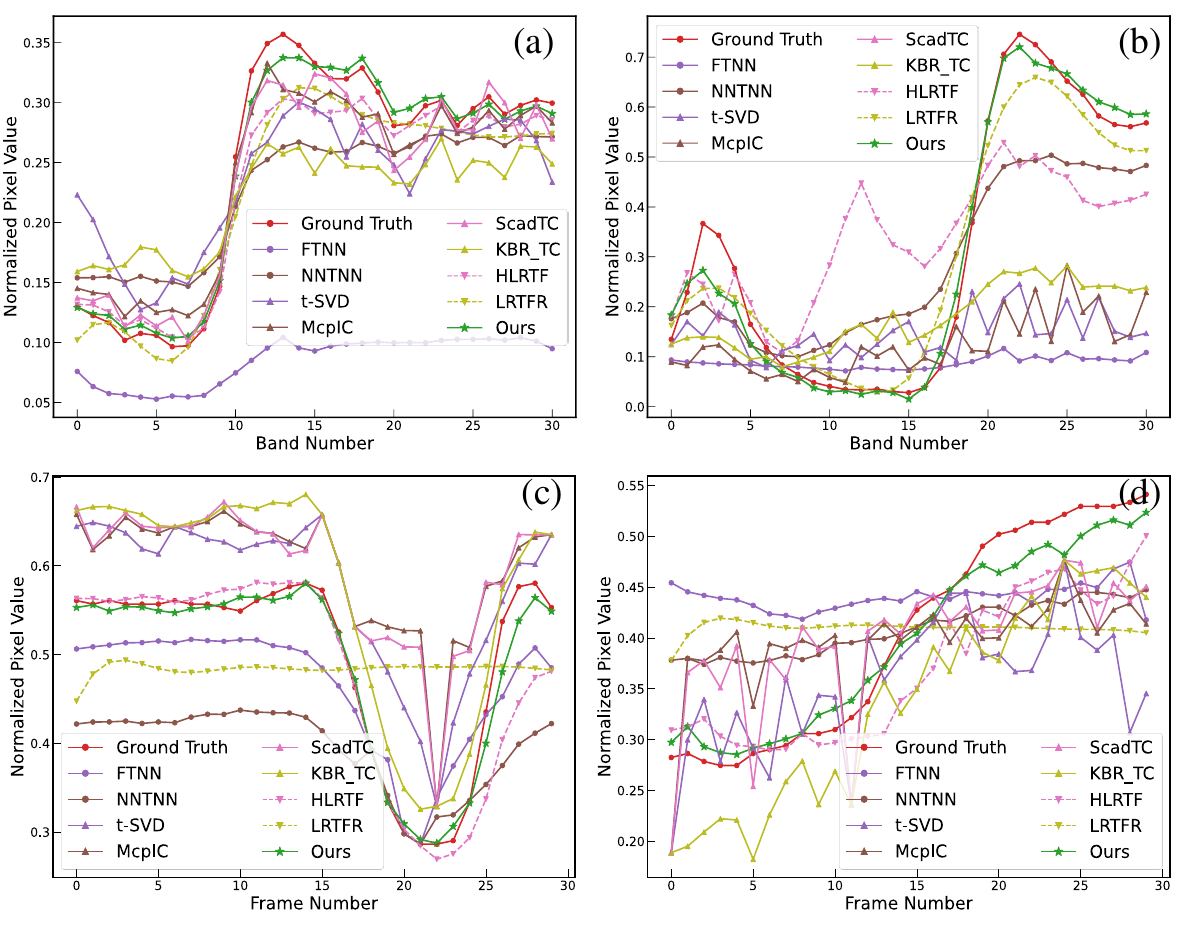}
\caption{The pixel values along the temporal dimensional (the same location of each frame) of the recovered (a)(b) hyperspectral image \emph{Balloons} and (c)(d) grayscale video \emph{Suzie}.}
\label{fig4}
\end{figure}

\begin{figure}[htbp]
\centering
\includegraphics[width=0.5\textwidth]{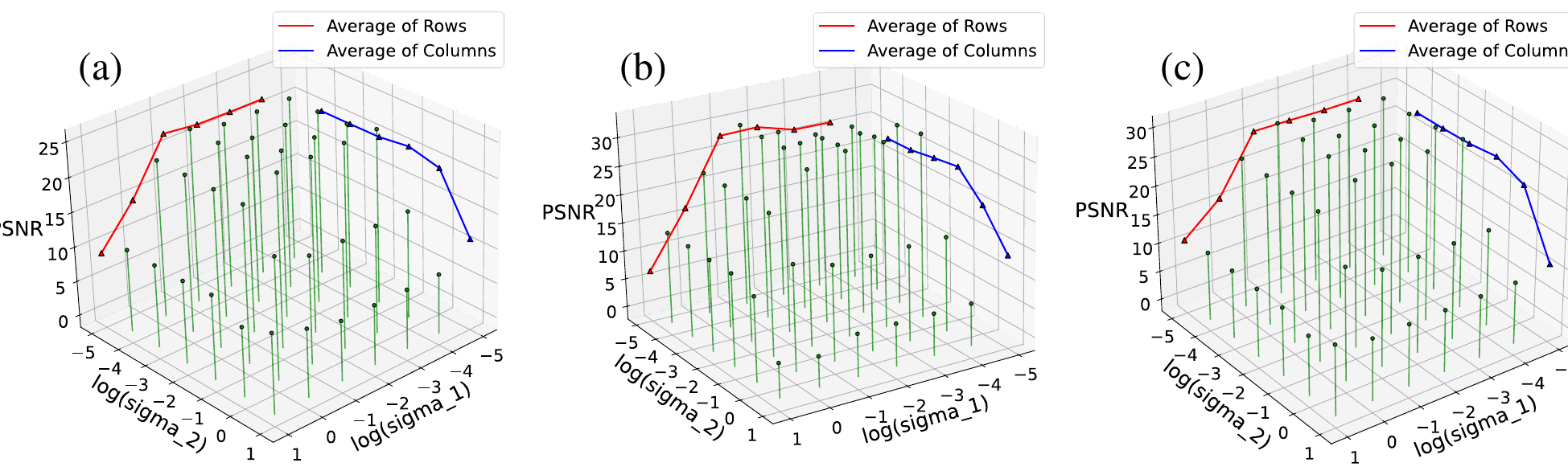}
\caption{Effect of different smoothing intensities on the reconstruction quality for (a) color image \emph{House}, (b) hyperspectral image \emph{Toy} and (c) grayscale video \emph{Suzie}.}
\label{fig5}
\end{figure}

\begin{figure}[htbp]
\centering
\includegraphics[width=0.5\textwidth]{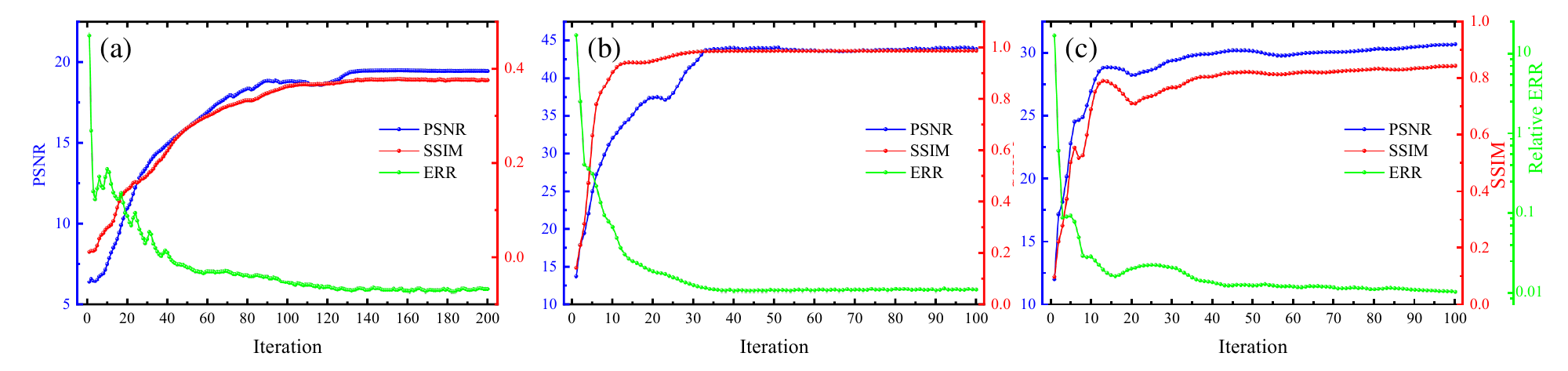}
\caption{Evolutions of PSNR, SSIM, and relative change rate during iteration for (a) color image \emph{Baboon}, (b) hyperspectral image \emph{Balloons} and (c) grayscale video \emph{Suzie}.}
\label{fig6}
\end{figure}
Figure \ref{fig3} illustrates the PSNR and SSIM for all bands and all frames after reconstruction in the Balloons and Suzie datasets. The quality of each frame reconstructed using the proposed algorithm is predominantly superior to, or at least comparable with, that of other algorithms. Moreover, the reconstruction quality exhibits minimal fluctuation, with no cases of complete reconstruction failure for any frame. This demonstrates the robustness and reliability of the proposed method across diverse data sets and conditions. 

Figure \ref{fig4} depicts the grayscale variations at a fixed pixel location. The closer the variation curve is to the actual values, the better the data reconstruction. Compared to the data recovered by other algorithms, the data reconstructed by the proposed method is closest to the fully sampled data.

Figure \ref{fig5} illustrates the effect of different local smoothing and non-local regularization strengths on the reconstruction results. The input data for the smoothing process is the reconstructed data considering only the low-rank prior. Different local smoothing intensities $\sigma_1$ and non-local smoothing intensity $\sigma_2$ are traversed to determine the optimal smoothing intensity by comparing the smoothed results. The results show that the reconstruction performance of the three datasets is best when both $\sigma_1$ and $\sigma_2$ are set to $0.1$.

For the proposed tensor reconstruction model, since the objective function is convex, the convergence of the ADMM-based algorithm is theoretically guaranteed. Figure \ref{fig6} demonstrates the PSNR, SSIM, and relative change rate changes during iterations. As shown in Figures 6(b) and 6(c), for hyperspectral images and grayscale videos, PSNR and SSIM gradually increase with more iterations and stabilize when the number of iterations exceeds 60. At the same time, the relative change rate approaches zero around 100 iterations. For color images, due to the limited information from only three channels, more iterations are required for convergence, as shown in Figure 6(a).

\subsection{Ablation Study}
In this work, we conducted three ablation experiments to evaluate the effects of removing individual priors: local smoothing, non-local similarity, and both simultaneously. The results of these experiments are presented in Table \ref{tab4}. The reconstruction performance declines when either the local smoothing prior or the non-local similarity prior is removed individually. The degradation becomes even more pronounced when both priors are removed, leaving only the low-rank prior. These findings highlight that the three priors contribute to the reconstruction process from different perspectives and work in a complementary and symbiotic manner.

\begin{table}[htbp]
\caption{Results of ablation study}
\begin{center}
\begin{adjustbox}{width=0.5\textwidth}
\begin{tabular}{ccccccccc}
\hline
&\multicolumn{2}{c}{\textbf{Multiple Priors}}&\multicolumn{2}{c}{\textbf{Remove CNN}}&\multicolumn{2}{c}{\textbf{Remove BM3D}}&\multicolumn{2}{c}{\textbf{Remove CNN and BM3D}} \\
\cline{2-9} 
 & \textit{PSNR} & \textit{SSIM} & \textit{PSNR}$\downarrow$ & \textit{SSIM}$\downarrow$ & \textit{PSNR}$\downarrow$ & \textit{SSIM}$\downarrow$ & \textit{PSNR}$\downarrow$ & \textit{SSIM}$\downarrow$\\
\hline
house&28.53&0.82&24.95 &0.75 &24.95 &0.74 &23.63 &0.58 \\
toy&36.58&0.96&34.74&0.946&34.24&0.93&33.25&0.89\\
suzie&30.87&0.84&29.28&0.83&29.22&0.80&28.28&0.75\\
\hline
\end{tabular}
\label{tab4}
\end{adjustbox}
\end{center}
\end{table}

\section{Conclusions}
We propose a tensor reconstruction method that integrates multiple priors to fully exploit prior information. The method enforces the low-rank property of the reconstructed data using learnable tensor decomposition, applies local smoothing via a pre-trained CNN, and incorporates BM3D regularization for non-local smoothing. The proposed ADMM-based algorithm solves the optimization problem by decomposing it into three subproblems, enabling efficient resolution. Comparative experiments across three datasets demonstrate that the proposed algorithm achieves superior reconstruction quality and visual performance, significantly outperforming other methods, particularly in extreme cases with sampling rates ranging from 1\% \text{to} 5\%.

\section{Data availability}
Data are available on request from the authors.

\section{Declaration of interests}
The authors declare that they have no known competing financial interests or personal relationships that could have appeared to influence the work reported in this paper.

\bibliographystyle{IEEEbib}
\bibliography{references}

\vspace{12pt}
\end{document}